# Ultrafast quantum-well photodetectors operating at 10μm with flat frequency response up to 70GHz at room temperature

M. Hakl[1†], Q.Y. Lin[1†], S. Lepillet[1], M. Billet[1‡], J-F. Lampin[1], S. Pirotta[2], R. Colombelli[2], W. J. Wan[3], J. C. Cao[3], H. Li[3], E. Peytavit[1], and S. Barbieri[1*]

[1]Institute of Electronics, Microelectronics and Nanotechnology, Univ. Lille, ISEN, CNRS, UMR 8520, 59652 Villeneuve d'Ascq, France
[2] Centre de Nanosciences et de Nanotechnologies (C2N), CNRS UMR 9001, Université Paris-Saclay, 91120 Palaiseau, France
[3]Key Laboratory of Terahertz Solid State Technology, Chinese Academy of Sciences, Shanghai, 200050, China

*Corresponding author: stefano.barbieri@univ-lille.fr
† These authors contributed equally to this work
‡ Present address: Photonics Research Group, Department of Information Technology, Ghent University IMEC, Ghent B-9000, Belgium

## Abstract

III-V semiconductor mid-infrared photodetectors based on intersubband transitions hold a great potential for ultra-high-speed operation up to several hundreds of GHz. In this work we exploit a ~350nm-thick GaAs/Al$_{0.2}$Ga$_{0.8}$As multi-quantum-well heterostructure to demonstrate heterodyne detection at λ~10μm with a nearly flat frequency response up to 70GHz at room temperature, solely limited by the measurement system bandwidth. This is the broadest RF-bandwidth reported to date for a quantum-well mid-infrared photodetector. Responsivities of 0.15A/W and 1.5A/W are obtained at 300K and 77K respectively. To allow ultrafast operation and illumination at normal incidence, the detector consists of a 50Ω coplanar waveguide, monolithically integrated with a 2D-array of sub-wavelength patch antennas, electrically interconnected by suspended wires. With this device architecture we obtain a parasitic capacitance of ~30fF, corresponding to the static capacitance of the antennas, yielding a RC-limited 3dB cutoff frequency >150GHz at 300K, extracted with a small-signal equivalent circuit model. Using this model, we quantitively reproduce the detector frequency response and find intrinsic roll-off time constants as low as 1ps at room temperature.

# 1. Introduction

Thanks to their intrinsically short electron relaxation time, on the ps timescale, mid-infrared (MIR-3-12 μm) quantum-well infrared photodetectors (QWIP) based on III-V semiconductor materials were identified as ideal candidates for ultra-high-speed operation at the end of the 80s. Since then, several experiments have been carried out to determine their RF bandwidth using both pulsed mid-infrared excitation or heterodyne detection [1-7].

The exploitation of QWIPs as heterodyne receivers with IF bandwidth of tens of GHz is particularly attractive for a number of applications, including free-space communications, gas sensing and spectroscopy, atmospheric and space science, or military countermeasures [8-11]. Besides enabling the implementation of coherent detection schemes, another advantage brought by heterodyne detection is the possibility to operate QWIPs in the shot-noise regime, overcoming the noise contribution of the thermally activated dark current, which severely impacts the NEP of MIR QWIPs at high temperatures [12]. So far, the largest heterodyne detection bandwidths at room temperature are those obtained by Grant et al. using a 100-quantum wells (QWs) QWIP operating at a 10μm [6,7]. The device was processed in a 16μm-side square mesa, illuminated from a 45° polished substrate. At room temperature, a ~25GHz 3dB cutoff and ~10dB attenuation at 110GHz were demonstrated, clearly illustrating the high frequency potential of QWIPs.

In the context of high-speed QWIPs the possibility of coupling the detector element to an antenna opens interesting perspectives. Indeed, the antenna allows a reduction of the detector active volume without sacrificing the radiation collection area, thus avoiding a reduction of the quantum efficiency. Nano-antennas were first applied to MIR bolometers as a way to increase both their sensitivity and speed [13-15]. QWIPs based on arrayed patch antennas resonators (PARs) were first proposed in 2001 [16]. PARs are ideally suited for QWIPs as they allow illumination at normal incidence, which is clearly advantageous compared to facet illumination [6,7], while confining the electro-magnetic field inside a sub-wavelength volume [16-18]. Compared to QWIPs based on standard mesa geometry and of comparable collection area, this enables the realization of "thin" detectors (including a small number of quantum wells) with a higher detectivity while keeping a small capacitance, which is clearly relevant for high speed operation. QWIPs based on 2D arrays of PARs were recently demonstrated showing over one order of magnitude improvement in detectivity compared to a mesa reference QWIP [19-20]. However, in these works the potential in terms of high-speed operation was largely underexploited, with reported maximum heterodyne detection frequencies (no reported 3dB bandwidth) up to only 4 GHz, limited by a device design (parasitics) leading to a large capacitance [20]. In this work we have fully addressed this issue by demonstrating PARs-based QWIP detectors specifically designed for ultra-broadband operation. Thanks to this design we demonstrate experimentally that the detector capacitance can be reduced down to the (unavoidable) static capacitance of the antenna resonators. With these devices we demonstrate room temperature heterodyne detection at 10.3 μm with a nearly flat frequency response up to 70GHz (limited by the detection electronics), and state of the art responsivities of ~0.15A/W. To the best of our knowledge this represents the broadest experimental RF bandwidth reported to date for a QWIP detector [6,7,21], and extends by over 65 GHz the results presented in Ref. [20]. Moreover, we develop a small-signal equivalent circuit model that can quantitively reproduce the observed device frequency response, which we find to be strongly dependent on bias and temperature. From this model we extract an RC limited 3dB cutoff of ~150GHz at 300K, and an intrinsic roll-off time constant ≲1ps, providing the first experimental evidence that QWIP detectors can indeed reach RF-bandwidths limited by electron capture on the ps timescale at room temperature. These results pave the way to the development of ultrafast MIR optoelectronics.

## 2. Device design and fabrication

The structure is grown by molecular beam epitaxy (MBE) on a semi-insulating GaAs substrate: 100nm-thick lattice-matched $Ga_{0.51}In_{0.49}P$ etch-stop layer followed by an $Al_{0.2}Ga_{0.8}As$/GaAs heterostructure. The PAR active region consists of seven, 6.5nm GaAs quantum wells (QWs) with a central, 5.3nm-thick region, n-doped at a level of $6.7 \times 10^{17}$ cm$^{-3}$. The wells width is chosen to obtain a bound-to-bound transition energy of ~120meV. The QWs are separated by 40nm barriers, and the active region is sandwiched between 50nm and 100nm-thick top and bottom n-doped contact layers with concentrations of $3 \times 10^{18}$ cm$^{-3}$ and $4 \times 10^{18}$ cm$^{-3}$ respectively.

In Fig. 1(a)(b), we present, the SEM images of the fabricated detector. It consists of a 5x5 periodic array of square PARs of side $s$=1.85μm and period $p$=3.9μm, sitting on top of a Ti/Au (100/400nm) ground plane. As detailed in the next Section, the values of $p$ and $s$ are chosen to obtain a maximum PAR array absorption as close as possible to the intersubband transition energy. At the same time, to minimize the array capacitance, the number of patches is kept to the minimum needed to allow collecting 100% of the incident radiation (~20μm-diameter laser spotsize, see next Section).

Particular care was taken in the detector microwave design, aimed at reducing the effect of parasitic capacitances brought by electrical connections and contact pads, which limited RF operation up to a few GHz in Ref. [20]. As shown in Fig.1(a),(b) this is achieved by

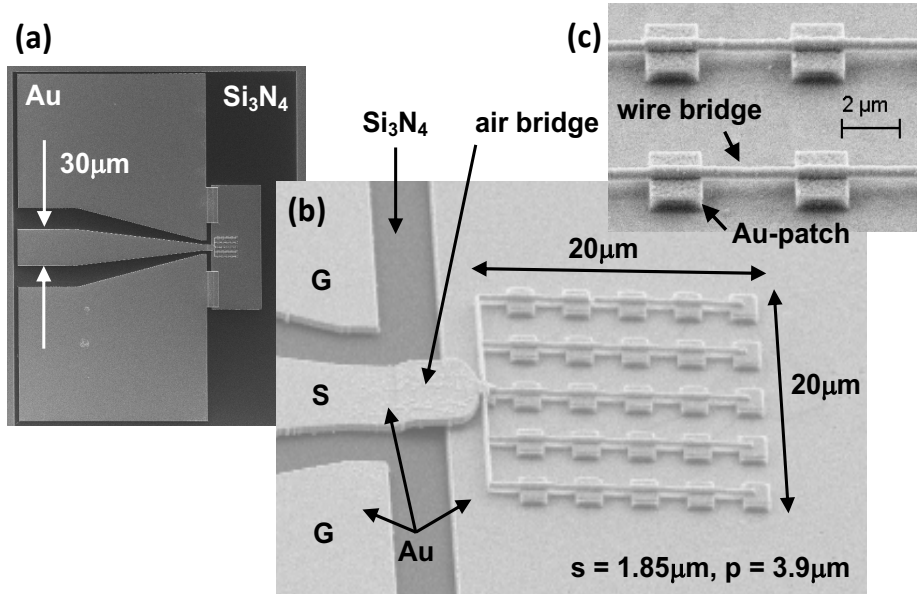

**Fig. 1 (a)** SEM image of the 5x5 PAR array with an integrated coplanar waveguide. **(b)** Close-up on panel (a) showing the full 5x5 PAR array used in the experiment ($s$=1.85μm; $p$=3.9μm), and the air bridge connecting the coplanar waveguide to the array. **(c)** Individual resonators incorporating the multi-QW structure are connected by suspended Au wires of ~150nm diameter (an array with $p$ = 5μm is shown in the panel).

connecting the 2D array to the central electrode of a 50Ω, tapered coplanar waveguide through an air bridge. Besides minimizing the parasitic capacitance, this solution is ideal for on-wafer testing by means of a 67GHz microwave coplanar probe. Finally, individual resonators are connected by suspended gold wires (Fig.1(b)(c)): compared to keeping the semiconductor beneath [20], this solution allows minimizing the wires capacitance, while simultaneously eliminating the current flow outside the resonators, therefore reducing the dark current. As a result of our design, as shown in Section 4, we find that the detector capacitance is essentially coincident with the static capacitance of the PARs alone, of approximately 30fF.

The fabrication of the PARs begins with the realization of a buried metal layer serving as electromagnetic ground plane and bottom Schottky contact metallization (we have deliberately chosen Schottky rather than ohmic contacts to avoid the risk of metal diffusion in the QWIP active region due to high temperature annealing, potentially leading to high MIR losses). This is obtained by transferring the epi-layers onto a 2"-diameter high-resistivity (>5 k$\Omega$.cm) silicon wafer using a Au–Au thermo- compression bonding technique detailed in [22], followed by the wet etching of the GaAs substrate and the etch-stop layer. Next, the Ti/Au (8nm/300nm) top Schottky contact metallization is realized through e-beam lithography, followed by e-beam evaporation and lift-off. The epi-layers are subsequently ICP etched using the top metal layer as etch-mask. The ground metal layer is finally dry-etched by an Ar$^+$ ion-beam around the PARs array down to the silicon substrate. A 100-nm-thick $Si_3N_4$ layer is then deposited on the silicon by plasma enhanced chemical vapor deposition (Fig.1(a)(b))

To electrically connect the patches together, suspended ~150-nm-width Ti/Au (20nm/600nm) wire-bridges are finally fabricated by a two-step e-beam lithography process. To this end a first resist layer is used as support after deposition, e-beam lithography and reflow, followed by a second one to define the wires by standard lift-off process. The same process is used to realize the air-bridge connecting the 2D array to the 50$\Omega$ coplanar line. The latter is deposited on the $Si_3N_4$: this avoids any leakage currents between the electrodes of the coplanar waveguide coming from the silicon.

## 3. Spectral and *dc* electrical characterisation

In Fig.2 we report the results of the infrared spectral characterization of the PARs array. Fig. 2(a) shows the absorption spectra at 300K, corresponding to the fraction of the incident power absorbed by the QWIP detector with two polarizations of the incident light: orthogonal (black) and parallel (red) to the wire bridges [19,20]. The absorption is defined as 1-$R(\omega)$, where $R(\omega)$ is the reflectivity spectrum obtained through FTIR micro-reflectivity measurements. At the cavity resonance for the orthogonal polarization (116meV – 10.7$\mu$m) we find that, 1-$R(\omega)$ = 0.9, i.e. 90% of the incident photons are absorbed. Indeed, the period *p*=3.9$\mu$m, was selected to operate the QWIP as close as possible to the critical coupling regime, compatibly with the targeted intersubband transition energy [17]. In this condition, the single PAR collection area at the resonant frequency is given by 0.9x$p^2$, yielding a total collection area of ~340μm² (=(18.5μm)²) for the PAR array. As shown in the insets of Fig.2(a), for the parallel polarization, the spatial distribution of the cavity mode is modified by the presence of the wire bridges, This yields a blue shift of the cavity resonance, as well as a reduced integrated absorption.

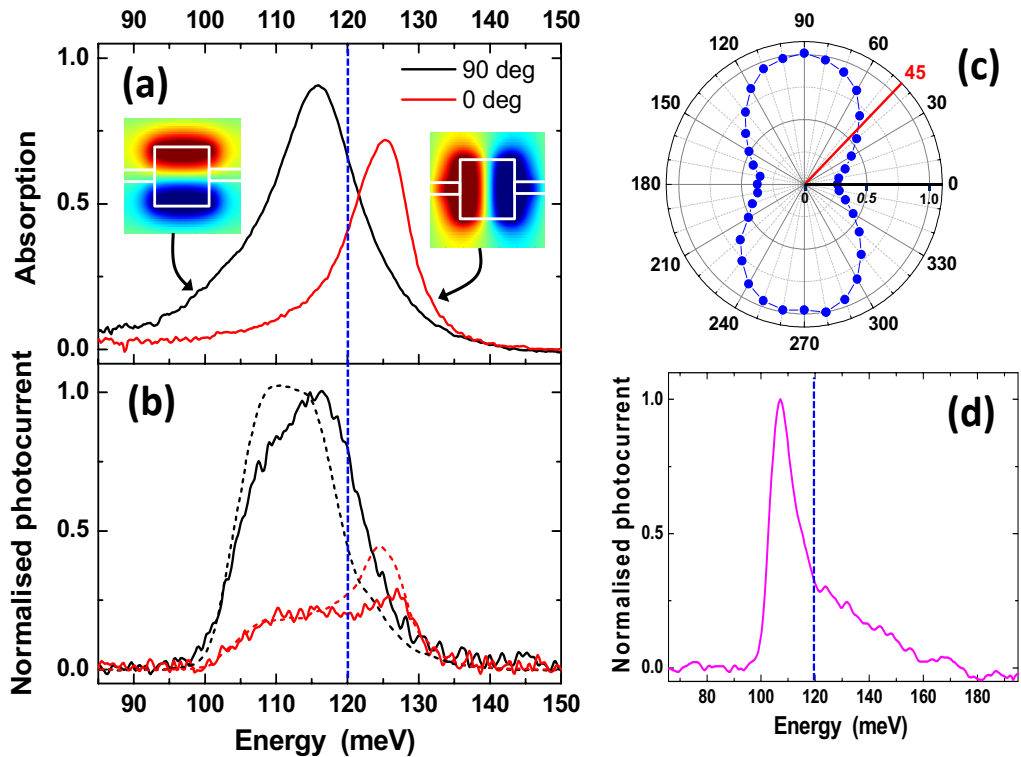

**Fig. 2. (a)** Absorption spectra of the PARs 2D array measured at 300K in two orthogonal polarizations: perpendicular (black) and parallel (red) to the wire bridges (spectra recorded at 77K, not shown, are virtually identical). The measurements are performed using a MIR microscope connected to the spectrometer. Insets. PAR fundamental modes in the two polarizations: computed 2D spatial profiles of the electric field component perpendicular to the surface (blue – positive; red- negative). Plots were obtained using a commercial FDTD solver. **(b)**. Photocurrent spectra measured at 77K in the two orthogonal polarizations (solid lines). Both spectra are normalized to the peak of the photocurrent spectrum at 90°. Dashed lines: spectra obtained by multi plying the spectrum of panel (d) by the absorption spectra of panel (a). **(c)** Normalized photocurrent *vs* polarization angle, measured at 300K, with a quantum cascade laser emitting at 10.3μm (120meV – dashed blue lines in panels (a), (b), and (d)). The red line indicates the polarization angle (45°) used for the measurements displayed in Fig.3 and Fig.4. **(d)** Photocurrent spectrum measured at 77K ($V_{bias}$ = 0.25V) with the QWIP processed in a mesa geometry.

In Fig.2(d) we report the measured photocurrent spectrum at 77K, obtained by FTIR spectroscopy with the QWIP structure processed in a mesa geometry, therefore showing the effect of the bare intersubband transition. We find a peak at 107meV, in good agreement with the expected bound-to-bound transition energy. By multiplying this spectrum by the cavity absorptions in Fig.2(a) we obtain the dashed spectra shown in panel (b), in good agreement with the QWIP detector photocurrent spectra measured a 77K, represented by the solid lines. This indicates that the PARs array absorption is dominated by absorption in the metal and contact layers (see Supporting Information for more details). From the black solid line we find that the QWIP operates in the ~10-12μm range, with a maximum response at ~10.8μm (115meV).

In Fig.3(a),(b) we report the dark current and *dc* photocurrent *vs* bias characteristics at 77K and 300K, obtained by illuminating the QWIP with a 10.3μm (120meV) DFB quantum cascade laser (QCL), polarized at 45° with respect to the wire bridges (the full polarization dependence at 77K is reported in Fig.2(c)). For these measurements, the collimated beam from the QCL was focused on the detector using an AR coated aspheric lens (NA = 0.56; 5mm focal length). At 10.3μm we measured a waist diameter of 20μm using a knife-edge technique, i.e. approximately equal to the side of the 5x5 PAR array collection area (18.5μm ~ $\sqrt{340 \mu m^2}$ ). Therefore, for the rest of this work, we assume that all the QCL power, measured after the lens, is incident on the QWIP. This corresponds to the power values reported in Fig.3(a)(b).

As expected, at 300K the dark current dominates the photocurrent for all power levels. At 77K the situation is reversed, showing that at this temperature the QWIP can be potentially operated in the photon-noise regime with only a few mW of incident power [2]. At 77K and 3.5-4V (Fig.3(a)) we observe a pronounced saturation of the photocurrent, that we attribute to negative differential drift velocity, resulting from intervalley scattering [23]. Saturation fields in the 10-20kV/cm range have been found in previous works. Here, at 3.9V (Fig.3(a)) the average electric field is ~100kV/cm, indicating that a large fraction of the applied bias drops on the Schottky contacts.

The photocurrent and responsivity as a function of incident power at 77K and 300K are reported in Fig.3(c), respectively at 3.4V and 2.5V. Responsivities are corrected by the polarization factor (Fig.2(c)), and their value corresponds to the situation where the incident field is polarized orthogonally to the wires, which is the ideal condition to operate the QWIP.

At low power we obtain responsivities R = 1.5A/W and 0.15A/W at 77K and 300K. From the reflectivity spectrum of Fig.2(a) (black line) and assuming an intersubband transition energy centered at 107meV with a FWHM of 10% (see the spectrum of Fig.1(d)), we find that the responsivity measured at 77K at high bias and low optical power, is compatible with a photoconductive gain $g = \tau_c/\tau_{tr} \cong 2.5$, where $\tau_c$ and $\tau_{tr}$ are respectively the electron's capture and transit time (see Supporting Information) [24]. The decrease of responsivity at 300K is attributed to a decrease of the drift velocity and capture time (see Table 1). Finally, by increasing the power we observe a clear decrease of responsivity at 77K. This is attributed to the presence of a series resistance provided by the Schottky contacts ($R_S$ in the circuit of Fig.3(d), see next Section and Table 1). As a consequence, for a given applied bias, the decrease of the detector photoresistance ($R_0$ in the circuit of Fig.3(d)) with increasing incident power produces a progressive lowering of the electric field across the QWIP active region [25] At room temperature $R_S$ is instead negligible (see Table 1), therefore the saturation effect is much less pronounced.

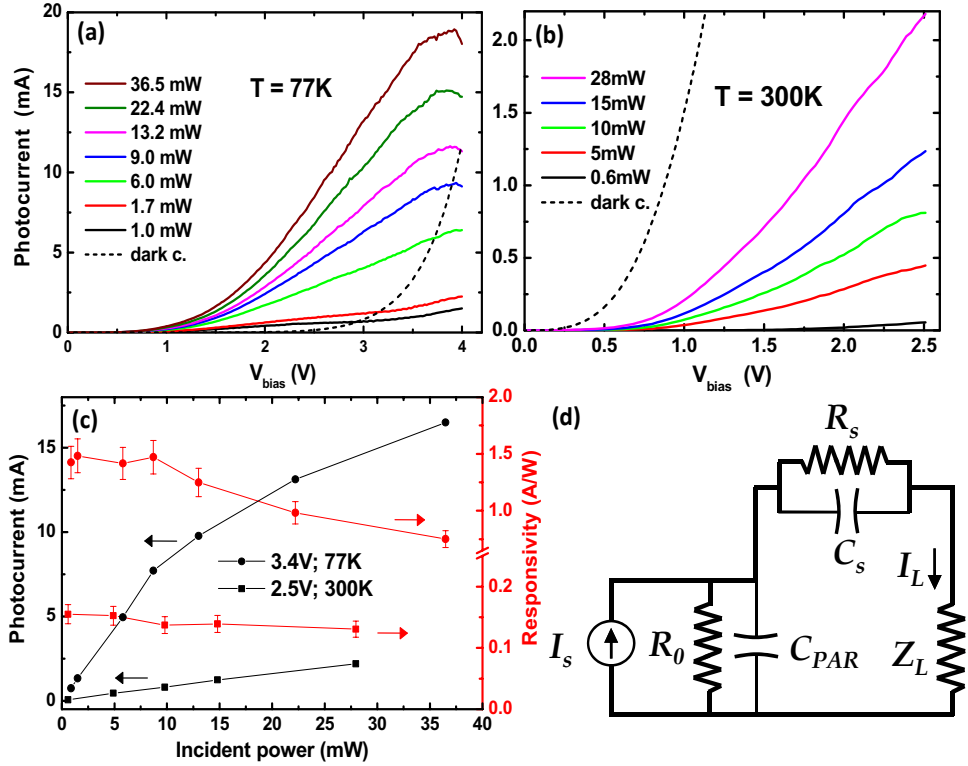

**Fig. 3**. Photocurrent *vs* applied bias at **(a)** 77K and **(b)** 300K for different incident QCL powers. The dark current I/V characteristics are shown in dashed. **(c)** Photocurrents (black dots) and responsivities (red dots) *vs* power, measured at 2.5V, 300K (squares) and 3.4V, 77K (circles). **(d)** Small signal; equivalent circuit of the QWIP detector (see text). $C_{PAR}$ ~30fF and $C_s$ ~1pF are respectively the 2D PARs array and Schottky contact capacitances. $R_0$ and $R_s$, (see Table 1) are respectively the *dc* internal photoresistance of the PARs array, and the leakage resistance of the Schottky contact biased in reverse breakdown, both under illumination (the forward biased Schottky junction is considered as a short circuit). $Z_L$ ~50Ω is the load impedance seen by the QWIP.

## 4. Heterodyne mixing and frequency response

In Fig. 4 we report the heterodyne frequency response (FR) of the QWIP in the 10MHz-67GHz range, measured at 77K and 300K, at low and high applied biases. To record these spectra, we used a 67GHz-bandwidth cryogenic probe, positioned at the edge of the coplanar waveguide shown in Fig.1(a). The photodetector was connected to a wideband bias-T and simultaneously illuminated by two 10.3μm-wavelength DFB QCLs driven with ultra-low noise (~300pA/Hz$^{1/2}$) current generators (see Supporting Information for a schematic of the experimental setup). The current of one QCL was kept constant while the current and temperature of the second one were fine-tuned in order to sweep the heterodyne frequency in the range 0-67GHz. The powers incident on the QWIP from the two QCLs are $P_1$ = 27.5mW and $P_2$ = 6mW (33.5mW total). The spectra of Fig.4 correspond to the intensities of the heterodyne beat signals recorded with a spectrum analyzer (SA) set in max-hold trace mode. The traces are corrected by (i) the propagation losses from the QWIP to the SA measured with a vector network analyzer (VNA), and (ii) the power changes (2dB max) of one QCL due to temperature/current tuning (see Supporting Information).

The top two traces in Fig.4 show the detector FR at high bias, i.e. 3.4V(77K) and 2.5V(300K). From Fig.2(c), the corresponding responsivities are 0.75A/W and 0.13A/W. At 77K we find a monotonic decrease with frequency, with a 3dB-cutoff frequency of ~30GHz, while at 300K the response is much flatter, with a ~2dB increase from 0 to ~40GHz, followed by a 3dB drop at ~67GHz.

At low biases the shape of the FR is rather different. As shown by the two bottom traces, recorded at 1.1V(77K) and 0.9V(300K), the FR is virtually flat up to 67GHz, except at low frequencies where we observe a pronounced drop below ~5GHz(77K) and ~10GHz(300K).

To gain insight in the behaviour of the QWIP, we used a VNA analyzer to derive the device impedance *vs* frequency in the operating conditions corresponding to the spectra of Fig.4. We find that at low bias (1.1V, and 0.9V spectra in Fig.4) the detector's RF impedance can be well reproduced using the simple small-signal circuit displayed in Fig.3(d) (see Supporting Information for the for the complete derivation of the circuit parameters) [26]. Here $R_0$, $R_s$, and

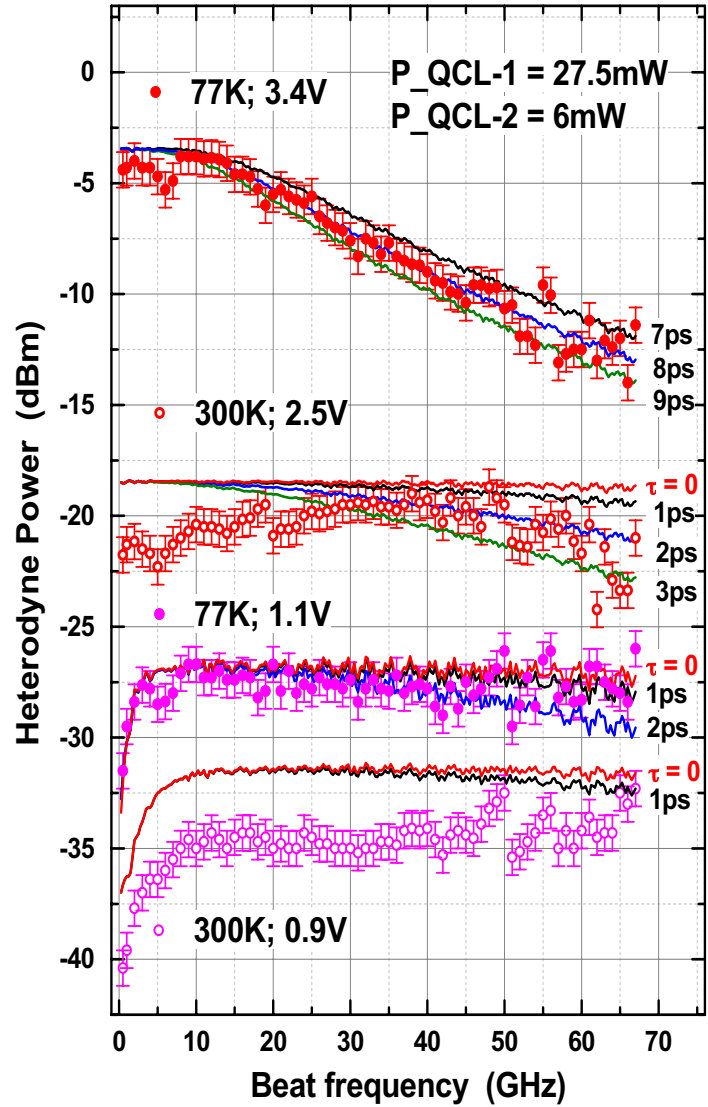

**Fig.4** QWIP detector FRs at different temperatures and biases (dotted curves). The powers incident on the QWIP from the two QCLs are $P_1$ = 27.5mW and $P_2$ = 6mW (33.5mW total). The spectra are corrected by the attenuation from the QWIP to the SA, measured with a VNA analyzer. The solid lines correspond to fits obtained using the small-signal circuit model of Fig.3(d) for different carrier's lifetimes (see main text).

$Z_L$~50Ω represent (i) the *dc* internal photoresistance of the 5x5 PAR array under illumination, (ii) the resistance of the reverse biased Schottky contact under illumination (the forward biased Schottky junction is considered as a short circuit), and (iii) the measured load impedance seen by the QWIP detector. $C_{PAR}$ ~30fF and $C_s$ ~1pF, are respectively the 2D PARs array and Schottky contact capacitance. The former corresponds to the computed static capacitance of the PARs array

The current source in the circuit represents the photocurrent generated in the patch array oscillating at the beat frequency $\omega_b$. It is given by:

$$I_s = \frac{m}{\sqrt{1+(\omega_b \tau)^2}} I_{ph} \frac{R_d+R_s}{R_d} = \frac{m}{\sqrt{1+(\omega_b \tau)^2}} I_0, \quad (1)$$

where *m* is a modulation index given by $m = \frac{2\sqrt{P_1 \times P_2}}{P_1+P_2} = 0.77$, $I_{ph}$ is the *dc* (i.e. average) measured photocurrent, $R_d$ is the PARs array dark resistance, and $I_0 = I_{ph}(R_d + R_s)/R_d$. The term under the square root at the denominator takes into account the frequency roll-off of the intrinsic transport mechanism, with τ approximating the carriers capture or transit time [24,26]. Concerning $R_d$, based on the fact that at 77K (300K) the QWIP current under illumination is dominated by the photocurrent (dark current) component (see Fig.3(a)(b)), we have made the following approximations: $R_d \approx R_0$ at 300K, and $R_d \gg R_s$ at 77K. These approximations allow *de facto* to eliminate the QWIP dark resistance as independent variable from the source term in Eq.(1) (see Supporting Information for more details)

The solid curves corresponding to the two bottom FRs in Fig.4 represent the power dissipated in the load: $P_L = \frac{1}{2}\mathcal{R}e[Z_L] \cdot |I_L|^2$. They are computed from the circuit of Fig.3(d) (with Eq.(1) for $I_s$) using (i) $R_0$ =200Ω, $R_s$=350Ω, for the spectrum at 1.1V (77K) with $I_{ph}$ =0.49mA and τ =1- 2ps; and (ii) $R_0$ =75Ω, $R_s$=125Ω, for the spectrum at 0.9V, (300K) with $I_{ph}$ =0.14mA and τ ~1ps (see Table 1, 1st and 2nd column). Despite the fairly simple electrical model and the measurement uncertainties the agreement with the experimental FRs is very good, both in terms of absolute power and spectral shape. In particular the observed drop at low frequency reflects the additional conversion losses due to the heterodyne power dissipated in $R_S$ when $f_b \lesssim (2\pi R_s C_s)^{-1}$ (see Fig.3.d). At higher frequencies $R_S$ is instead shorted by $C_s$, thus eliminating the power loss in the contact resistance. In this case, from the small-signal circuit model, we have that:

$$I_L = I_S \frac{1}{1+R_L/R_0+i\omega_b R_L C_{PAR}} \quad (2)$$

yielding a parasitic roll-off time constant $R_L C_{PAR}/(1+R_L/R_0) \lesssim 1$ps (see Table 1). We also find (see Supporting Information) that, for the chosen PAR array size, the QWIP impedance at low bias is close to 50Ω for frequencies $\gtrsim$20GHz(300K) and 30GHz(77K), which is ideal for RF impedance matching.

| T(K) | 77 | 300 | 77 | 300 |
|---|---|---|---|---|
| $V_{bias}$(V) | 1.1 | 0.9 | 3.4 | 2.5 |
| $I_{ph}$(mA) | 0.49 | 0.14 | 15.2 | 2.2 |
| $I_0$(mA) | 0.49 | 0.38 | 15.2 | 2.2 |
| $R_0$(Ω) | 200 | 75 | 40 | 40 |
| $R_s$(Ω) | 350 | 125 | 20 | 0 |
| τ(ps) | 1.5 | $\lesssim$1 | 8 | 2.5 |
| $\tau_c$(ps) | 1 | $\lesssim$1 | 10 | 2.5 |
| $\tau_{tr}$(ps) | 25 | 90 | 8 | 14 |
| $v_d$ (x10$^6$cm/s) | 1.5 | $\gtrsim$0.4 | 4.6 | 2.6 |

**Table 1.** Measured photocurrents ($I_{ph}$), $I_0$, and small-signal circuit resistances under illumination ($R_0$, $R_s$), used to compute the solid lines in Fig.4 for different operating conditions (bias and temperature). The value of the roll-off time constant (τ in Eq.(1)) is the one yielding the best fit of the experimental data. The capture time ($\tau_c$) and transit time ($\tau_{tr}$) are obtained from τ and the photoconductive gain (see text). The corresponding drift velocity ($v_d$) is obtained from the ratio between the thickness of the QWIP active region (365nm) and $\tau_{tr}$.

At high biases the effect of $C_S$ is much less pronounced and the power drop at low frequencies disappears (Fig.4, top two spectra). From the small-signal circuit this can be explained by a reduction of $R_S$ due to the Schottky barrier becoming more transparent, therefore effectively shunting $C_s$ at low frequencies. As a result, the QWIP impedance does not display the strong increase at low frequency found at low biases (see Supporting Information). From the small-signal circuit we find a good agreement with the measured FRs using (i) $R_0$ =40Ω, $R_s$=20Ω, for the spectrum at 3.4V (77K) with, τ ~ 8ps and $I_{ph}$ =15.2mA; and (ii) $R_0$ =40Ω, $R_s$=0Ω, for the spectrum at 2.5V, (300K) with τ ~2-3ps and $I_{ph}$ =2.2mA (see Table 1, 3[d] and 4[th] column). As a result, at 300K the QWIP is almost impedance matched to 50Ω at all frequencies. We note that our small signal circuit model does not explain the ~2dB increase in the FR from 0 to ~40GHz observed at 2.5V.

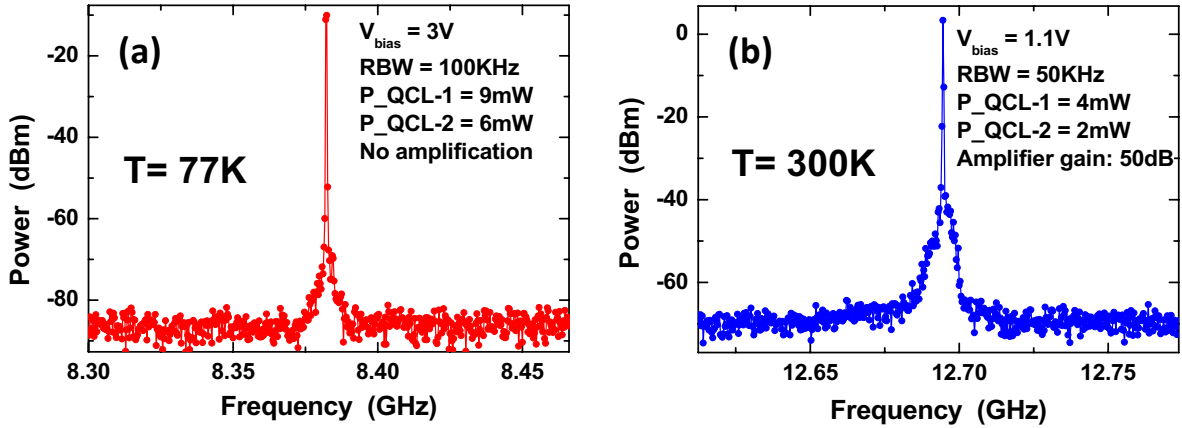

**Fig.5** Examples of single shot heterodyne beatnote spectra recorded (a) at 77K without amplification, and (b) at 300K with a low noise, narrow band amplifier of 50dB gain.

As shown above, thanks to the very small device capacitance, by fitting the measured FRs using the small signal circuit model we can extract the intrinsic detector response times, which, as shown by the solid curves in Fig.4, dependent on the operating conditions. From the values of $\tau$ and from the photoconductive gain derived from the responsivities, we can then obtain the values of $\tau_c$ and $\tau_{tr}$ shown in Table 1 (see Supporting Information): except at 77K under high bias, the QWIP intrinsic response time appears to be dominated by electron capture. We also find the expected decrease/increase of $\tau_{tr}$ with increasing bias/temperature [24]. Interpreting the dependence of $\tau_c$ on bias and temperature is beyond the scope of this work and will require more systematic measurements that are presently under way. At the same time, on this subject there appears to be a lack of experimental data in the literature [4].

In Fig.5 we report two examples of heterodyne beatnote spectra recorded in single shot at 77K, under an applied bias of 3V and without any amplification (Fig.5(a)), and at 300K, with an applied bias of 1.1V and with a narrow band amplifier of 50dB gain (Fig.5(b)). In the first case the noise floor is limited by the spectrum analyzer, while in the second spectrum the noise floor is determined by the amplifiers noise. We find instantaneous linewidths of ~100kHz, limited by the QCL thermal and current fluctuations. At 77K the RBW is set to 100kHz, yielding a SNR of ~77dB, while at 300K we find a SNR of 72dB with a RBW of 50kHz. Reducing further the RBW produces a decrease of the beatnote intensity because the RBW goes below the instantaneous heterodyne beatnote linewidth.

The dependence of the SNR of the heterodyne beatnote frequency, obtained with the spectrum analyzer without amplification ($P_1$ = 27.5mW and $P_2$ = 6mW), can be directly extracted from the spectra recorded in max-hold trace mode (see Supporting Information). At 30GHz and 60GHz, with a RBW of 3.5MHz, we obtain SNRs of 50dB and 35dB, and of 35dB and 25dB, respectively at 77K (3.4V) and 300K (2.5V).

## 5. Conclusions

Antenna-coupled QWIP detectors operating in the 10μm-12μm range are demonstrated, exhibiting a flat frequency response up to 67GHz at 77K and 300K. At 300K, from our experimental results and with the help of an equivalent circuit model, we find an RC-limited 3dB cutoff frequency >150GHz and ~ps intrinsic response times. These results are achieved thanks to a detector specifically designed for ultrafast operation, and provide the first experimental evidence that QWIPs can indeed reach RF-bandwidths limited by electron capture over timescales of ~1ps at room temperature.

We believe that the detectors demonstrated here, in combination with QCLs, will open up new perspectives in MIR photonics, namely by extending to the MIR range the possibilities offered by ultra-fast near-infrared optoelectronics, so far the only frequency range benefitting from the availability of ultrafast photodetectors. Envisaged applications are free space communications with data rates >10Gb/s, coherent multi-species gas sensing, high precision spectroscopy and metrology, astronomy, as well as to study real time dynamics on the 10ps time scale [8-11,27]. More specifically, on this last topic, we expect that broadband devices, such as those demonstrated in this work can shed new light on the ultrafast electron's dynamics in QWIPs.

A final intriguing perspective is the use of these structures as QCL-pumped photomixers for the generation of sub-THz radiation [28,29]. To this end we note that the actual responsivity of ~0.75A/W at 77K, 3.4V (high power – see Fig.3(c)) could be theoretically brought up to ~4A/W if the ISB transition peak and the PAR resonance were perfectly matched. In this case, from the top trace in Fig.4, we would expect microwave power levels in the mW range. Replacing the Schottky contacts with non-diffusive ohmic contacts (not to increase MIR photon absorption) should also reduce the saturation of the responsivity at high incident powers (Fig.2(c), 77K).

**Funding.** Nord-Pas de Calais Regional Council; Fonds Européens de Développement Régional. RENATECH (French Network of Major Technology Centres). CPER "Photonics for Society". French National Research Agency, and Direction Générale de l'Armement (project HISPANID). French National Research Agency (project IRENA). European Union FET-Open Grant MIR-BOSE (737017) and the European Research Council (IDEASERC) ("GEM") (306661). National Natural Science Foundation of China (61875220); "From 0 to 1" Innovation Program of the Chinese Academy of Sciences (ZDBS-LY-JSC009).

**Supporting information**. Supporting Information Available: Computed responsivity; Circuit model; Heterodyne mixing experimental setup and frequency response spectra; QWIP impedance measurements; Determination of QWIP circuit parameters; Evaluation of carriers capture and transit time. This material is available free of charge via the Internet at http://pubs.acs.org.

**References**

1. C. G. Bethea, F. Levine, G. Hasnain, J. Walker, and R. J. Malik, High-speed measurement of the response time of a GaAs/Al$_x$Ga$_{1-x}$As multiquantum-well long-wavelength infrared detector J. Appl. Phys. **66**, 963 (1989)
2. H. C. Liu, J. Li, E. R. Brown, K. A. McIntosh, K. B. Nichols, and M. J. Manfra, Quantum Well Intersubband Heterodyne Infrared Detection Up to 82 GHz, Appl. Phys. Lett. **67**, 1594 (1995)
3. S. Ehret, H. Schneider, J. Fleissner, P. Koidl, Ultrafast intersubband photocurrent response in quantum-well infrared photodetectors, Appl. Phys. Lett. **71**, 641 (1997).
4. S. Steinkogler, H. Schneider, R. Rehm, M. Walther, P. Koidl, P. Grant, R. Dudek, H.C. Liu, Time-resolved electron transport studies on InGaAs/GaAs-QWIPs, Infr. Phys. Technol. **44**, 355 (2003)
5. S. Steinkogler, H. Schneider, M. Walther, P. Koidl, Appl. Phys. Lett. Determination of the electron capture time in quantum-well infrared photodetectors using time-resolved photocurrent measurement, 82, 3925 (2003)
6. P.D. Grant, R. Dudek, M. Buchanan, L. Wolfson, H.C. Liu, An ultrafast quantum well infrared photodetector, Infr. Phys. Technol. **47**, 144 (2005)
7. P. D. Grant, R. Dudek, M. Buchanan, and H. C. Liu, Room-temperature heterodyne detection up to 110GHz with a quantum well infrared photodetector, IEEE Photon. Technol. Lett. **18**, 2218 (2006)

8. X. Pang, O. Ozolins, R. Schatz, J. Storck, A. Udalcovs, J. R. Navarro, A. Kakkar, G. Maisons, M. Carras, G. Jacobsen, S. Popov, and S. Lourdudoss, Gigabit free-space multi-level signal transmission with a mid-infrared quantum cascade laser operating at room temperature, Opt. Lett. **42**, 3646 (2017).
9. J. J. Liu, B. L. Stann, K. K. Klett, P. S. Cho, and P. M. Pellegrino, Mid and Long-Wave Infrared Free-Space Optical Communication, Proc. SPIE **11133**, 1113302-1 (2020)
10. N. A. Macleod, F. Molero, and D. Weidmann, Broadband standoff detection of large molecules by mid-infrared active coherent laser spectrometry, Opt. Expr. **23**, 912 (2015)
11. D. Weidmann, W. J. Reburn, and K. M. Smith, Ground-based prototype quantum cascade laser heterodyne radiometer for atmospheric studies, Rev. Sci. Instr. Expr. **78**, 073017 (2007)
12. E. R. Brown, K. A. McIntosh, K. B. Nichols, F. W. Smith, and M. J. Manfra, $CO_2$-Laser Heterodyne Detection with GaAs/AlGaAs MQW Structures, in Quantum Well Intersubband Transition Physics and Devices, NATO ASI Series, 207-220 (Springer, 1994)
13. E. N. Grossman, J. E. Sauvageau, and D. G. McDonald, Lithographic spiral antennas at short wavelengths, Appl. Phys. Lett. **59**, 3225 (1991)
14. N. Chong and H. Ahmed, Antenna-coupled polycrystalline silicon air-bridge thermal detector for mid-infrared radiation, Appl. Phys. Lett. **71**, 1607 (1997)
15. F. J. González, B. Ilic, J. Alda, and G. D. Boreman, Antenna-Coupled Infrared Detectors for Imaging Applications, IEEE J. Sel. Top. Qauntum. Electron. **11**, 117 (2005)
16. A. Beck, and M. S. Mirotznik, Microstrip antenna coupling for quantum-well infrared photodetectors, Infr. Phys. Technol. **42**, 189 (2001)
17. D. Palaferri, Y. Todorov, A. Mottaghizadeh, G. Frucci, G. Biasiol and C. Sirtori, Ultra-subwavelength resonators for high temperature high performance quantum detectors, New J. Phys. **18**,113016 (2016)
18. J. Le Perchec, Y. Desieres, and R. Espiau de Lamaestre, Plasmon-based photosensors comprising a very thin semiconducting region, Appl. Phys. Lett. **94**, 181104 (2009)
19. Y. N. Chen, Y. Todorov, B. Askenazi, A. Vasanelli, G. Biasiol, R. Colombelli, and C. Sirtori Antenna-coupled microcavities for enhanced infrared photo-detection, Appl. Phys. Lett. **104**, 031113-1 (2014).
20. D. Palaferri, Y. Todorov, A. Bigioli, A. Mottaghizadeh, D. Gacemi, A. Calabrese, A. Vasanelli, L. Li, A.G. Davies, E. H. linfield, F. Kapsalidis, M. Beck, J. Faist and C. Sirtori, Room-temperature 9-μm wavelength photo-detectors and GHz-frequency heterodyne receivers, Nature **556**, 85 (2018).
21. E. Rodriguez, A. Mottaghizadeh, D. Gacemi, D. Palaferri, Z. Asghari, M. Jeannin, A. Vasanelli, A. Bigioli, Y. Todorov, M. Beck, J. Faist, Q. J. Wang, and C. Sirtori, Room temperature, wide-band Quantum Well Infrared Photodetector for microwave optical links at 4.9 μm wavelength, ACS Photonics **5**, 3689 (2018)
22. E. Peytavit, J.-F. Lampin, F. Hindle, C. Yang, and G. Mouret, Wide-band continuous-wave terahertz source with a vertically integrated photomixer, Appl. Phys. Lett., 95, 161102, (2009)
23. H. Schneider, C. Mermelstein, R. Rehm, C. Schönbein, A. Sa'ar, and M. Walther, Optically induced electric-field domains by bound-to-continuum transitions in n-type multiple quantum wells, Phys. Rev. B **57**, R15096(R) (1998)
24. Schneider, H. & Liu, H. C. in Quantum Well Infrared Photodetectors: Physics and Applications 72–75 (Springer, 2007).
25. M. Ershov, H. C. Liu, M. Buchanan, Z. R. Wasilewski, and V. Ryzhii Photoconductivity nonlinearity at high excitation power in quantum well infrared photodetectors Appl. Phys. Lett. **70**, 414 (1997)
26. P. D. Coleman, R. C. Eden, and J. N. Weaver, Mixing and Detection of Coherent Light, IEEE Trans. Electron Devices, 11, 488 (1964)
27. D.D.S. Hale, M. Bester, W. C. Danchi, W. Fitelson, S. Hoss, E. A. Lipman, J. D. Monnnier, P. G. Tuthill, and C. H. Townes, The Berkeley infrared spatial interferometer: a heterodyne stellar interferometer for the mid-infrared, Astrophysical Journal **537**, 998 (2000).
28. E. Peytavit S. Lepilliet, F. Hindle, C. Coinon, T. Akalin, G. Ducournau, G. Mouret, and J-F. Lampin, Milliwatt-level output power in the sub-terahertz range generated by photomixing in a GaAs photoconductor, Appl. Phys. Lett. **99**, 223508 (2011)
29. E. Peytavit, P. Latzel, F. Pavanello, G. Ducournau, and J.-F. Lampin, CW Source Based on Photomixing With Output Power Reaching 1.8 mW at 250 GHz, IEEE Electron. Dev. Lett., **34**, 1277 (2013)

# Supporting Information:
# Ultrafast quantum-well photodetectors operating at 10μm with flat frequency response up to 70GHz at room temperature

M. Hakl[1†], Q.Y. Lin[1†], S. Lepillet[1], M. Billet[1‡], J-F. Lampin[1], S. Pirotta[2], R. Colombelli[2], W. J. Wan[3], J. C. Cao[3], H. Li[3], E. Peytavit[1], and S. Barbieri[1*]

[1]Institute of Electronics, Microelectronics and Nanotechnology, Univ. Lille, ISEN, CNRS, UMR 8520, 59652 Villeneuve d'Ascq, France
[2] Centre de Nanosciences et de Nanotechnologies (C2N), CNRS UMR 9001, Université Paris-Sud, Université Paris-Saclay, 91120 Palaiseau, France
[3]Key Laboratory of Terahertz Solid State Technology, Chinese Academy of Sciences, Shanghai, 200050, China

*Corresponding author: stefano.barbieri@univ-lille.fr
† These authors contributed equally to this work
‡ Present address: Photonics Research Group, Department of Information Technology, Ghent University IMEC, Ghent B-9000, Belgium

## 1. Computed responsivity

The responsivity of the PAR array for an incident electromagnetic wave of frequency ω, polarized perpendicularly to the wire bridges, can be computed from [1]:

$$R_{QWIP}(\omega) = [1 - R(\omega)] \left[\frac{A_{isb}(\omega)}{A_{isb}(\omega) + Q_{PAR}^{-1}}\right] \left[\frac{eg}{\hbar\omega}\right], \quad (S1)$$

where $[1-R(\omega)]$ is the PAR array absorption spectrum shown in Fig2(a) of the main text, $g = \tau_c/\tau_{tr}$ is the photoconductive gain, $e$ is the electronic charge, and $A_{isb}(\omega)$ is the intersubband absorption coefficient of the PAR given by the following expression:

$$A_{isb}(\omega) = f_w \frac{\omega_p^2}{4\omega_{21}} \frac{\Gamma}{\hbar^2(\omega-\omega_{21})^2 + \Gamma^2/4}. \quad (S2)$$

Here $f_w$ = 0.088 is the overlap factor between the PAR mode and the quantum wells; $\hbar\omega_p$=27.5 meV is the intersubband plasma energy [2]; $\hbar\omega_{21} \approx 107$meV is the intersubband transition energy extracted from the photocurrent spectrum (see Fig.2(d) of the main text), and $\Gamma \approx 10.7$meV is the FWHM of the intersubband transition, which we assume to be approximately equal to 10% of $\hbar\omega_{21}$. $Q_{PAR} \approx 8$ in Eq.(S1) is the quality factor of the PAR array (i.e. excluding intersubband absorption), that we obtain from the FWHM of $[1-R(\omega)]$ (Fig.2(a) of the main text). This is a good approximation since, due to the spectral shift between $A_{isb}(\omega)$ and $[1-R(\omega)]$, absorption in the resonators should be dominated by ohmic losses (i.e. absorption in the metal and contact layers). Indeed, results from FDTD simulations performed on a single resonator yield an upper limit of ~15 for the PAR array Q-factor (without intersubband absorption). This is smaller than $1/A_{isb}(\omega_{12}) \approx 20$ obtained from Eq.(S2) showing that even without spectral shift, cavity absorption would be dominated by ohmic losses).

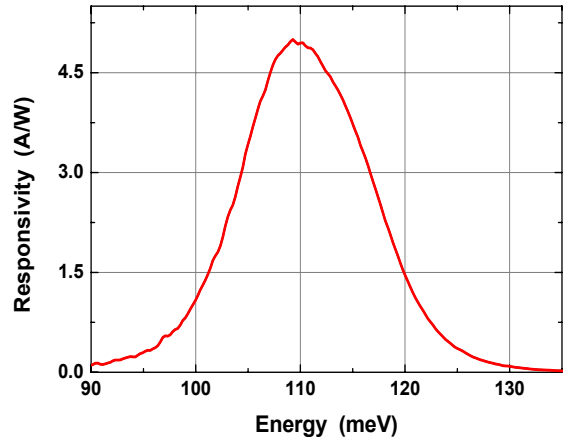

**Fig. S1 (a)** Computed responsivity spectrum from Eq.(S1) and (S2)

The responsivity $R_{QWIP}(\omega)$, obtained from Eq.(S1) and (S2) with $g = 2.5$, is shown in Fig.S1: for $\hbar\omega \approx 120$meV, corresponding to the QCL photon energy ($\lambda = 10.3\mu m$), we obtain $R_{QWIP}$=1.5A/W, in accordance with the measured experimental responsivity at 3.4V, 77K and low incident power (Fig.3(c) in the main text).

## 2. Heterodyne mixing experimental setup and frequency response spectra

The schematic of the experimental setup up is shown in Fig.S2. To minimize the linewidth of the heterodyne beatnote the QCLs were driven with low noise current drivers (Koheron, DRV110) with a current noise of 300pA/Hz$^{1/2}$ .

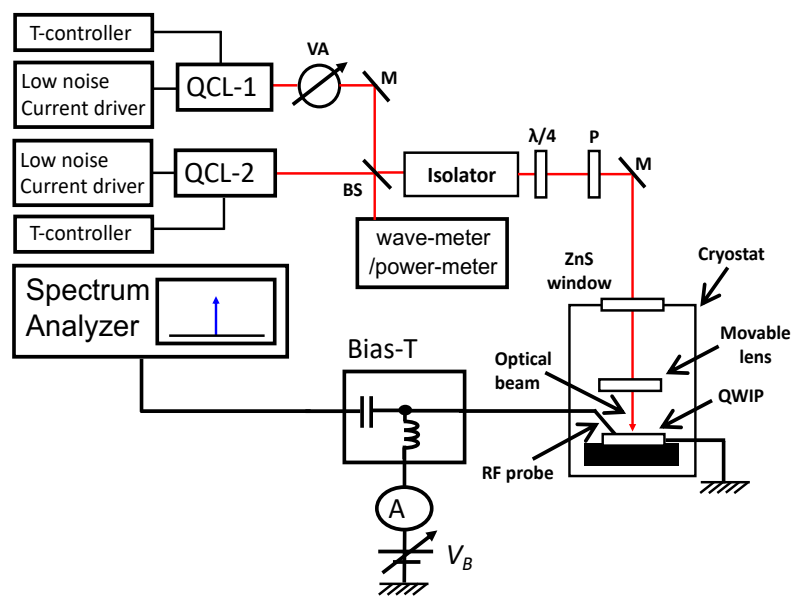

**Fig. S2** Schematic of the experimental setup used for the measurement of the QWIP FR. VA – variable attenuator; M – mirror; BS -beam splitter; λ/4 – quarter waveplate; P – polarizer.

Fig.S3 shows the procedure followed to extract the frequency response (FR) spectra displayed in Fig.4 of the main text. As an example we consider the FR at 300K and 2.5V: first, by sweeping the frequency of one QCL, we have recorded the heterodyne beat signal using the spectrum analyzer (SA) set in max-hold trace mode as described in the main text. The corresponding SA trace is displayed in

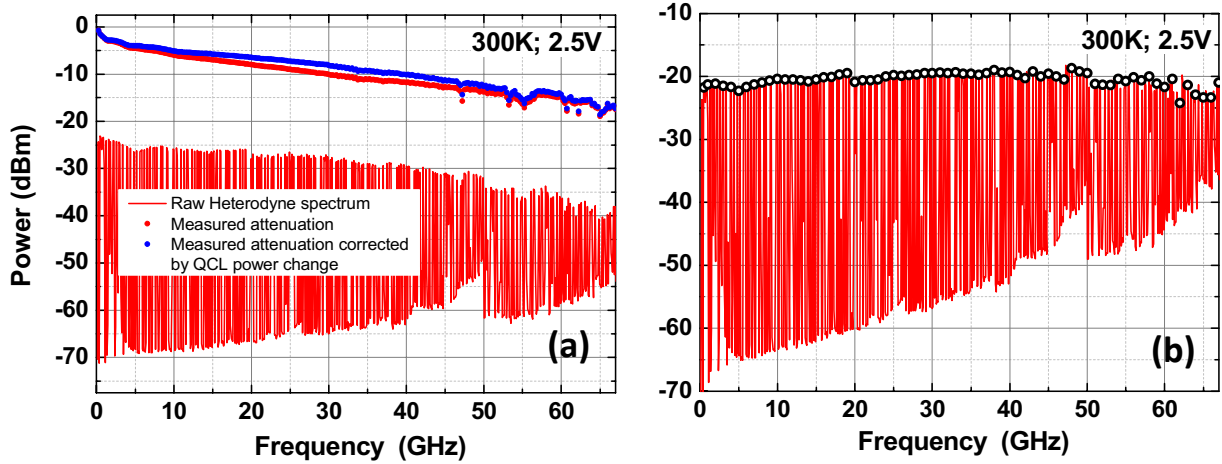

Fig. S3 (a) Example of extraction of the 300K; 2.5V heterodyne spectrum, collected with the SA set in max-hold trace mode. The raw heterodyne spectrum (red) is corrected by subtracting the attenuation (red dots) measured by injecting an RF signal into the SA, and compensated by the QCL power change (blue dots). (b) The resulting heterodyne beat spectrum corrected by the attenuation and QCL power change. The black circles correspond to the line peaks recorded every 500 MHz.

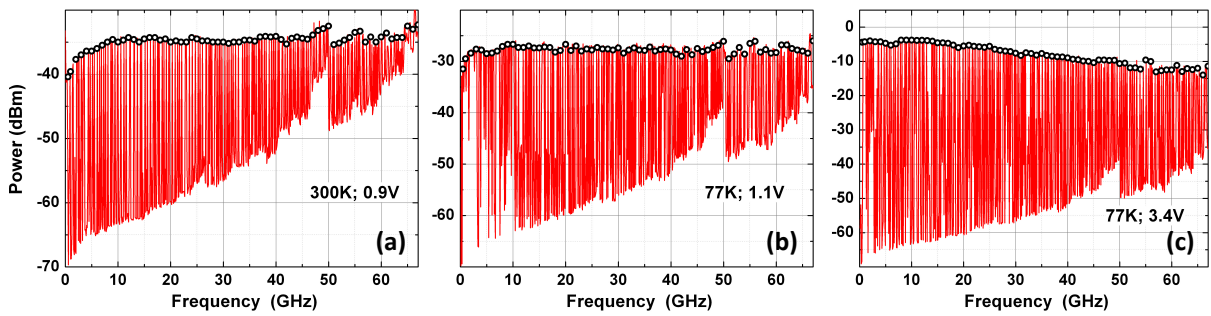

Fig. S4 (a) Heterodyne beat spectra corrected by the attenuation and QCL power change, following the same procedure used to obtain the spectrum of Fig.S3(b). The black dots are those displayed in Fig.4 of the main text (300K, 0.9V - 77K, 1,1V – 77K, 3.4V).

spectrum shown in Fig.S3(b). The black circles, corresponding to the line peaks recorded every 500MHz, are those displayed in Fig.4 of the main text. We note that the heterodyne spectra were recorded with a RBW of 3.5MHz. This is larger than the actual heterodyne beat linewidth, which was of the order of ~100kHz (see the main text), therefore guaranteeing that the intensity of the heterodyne beats is not reduced by filtering.

In Fig.S4 we report the other heterodyne spectra (corrected by losses) used to extract the FRs shown in Fig.4 of the main text.

## 3. ~~Small-signal~~ Circuit model

In an optical heterodyne experiment as described in this work, a *dc* biased photoconductor is illuminated by two laser beams of power $P_1$ and $P_2$ with a difference frequency $\omega_b$. The incident optical power on the photoconductor can be expressed as:

$$P(t) = P_1 + P_2 + 2\sqrt{P_1 P_2}\sin(\omega_b t). \qquad (S3)$$

The photocarrier density in the photoconductor follows the time variation of the incident power. By assuming that the photoconductor exhibits a linear I-V characteristic, it can be modeled as a time-dependent conductance, which can be written as:

$$G(t) = G_0 + G_1 \sin(\omega_b t + \varphi), \qquad (S4)$$

where $G_0$ and $G_1$ are respectively a *dc* and a dynamic conductance term. They are given by [3]:

$$G_0 = G_d + G_{ph}, \qquad (S5)$$

$$G_1 = \frac{m}{\sqrt{1+(\omega_b \tau)^2}} G_{ph}, \qquad (S6)$$

with

$$G_{ph} = \frac{I_{ph}}{V_{dc}}. \qquad (S7)$$

In Eq.(S6) and (S7), the term $G_d$ ($1/R_d$) is the dark conductance, while $G_{ph}$ ($1/R_{ph}$) is the internal photoconductance given by the ratio between the *dc* (i.e. average) conduction photocurrent, $I_{ph}$, generated by the two laser sources, and the *dc* bias voltage, $V_{dc}$ applied to the photoconductor. In the expression of $G_1$ the denominator reflects the frequency roll-off of the intrinsic recombination or transport mechanism, with τ approximating the carriers capture or transit time. The term *m* is a modulation index given by:

$$m = \frac{2\sqrt{P_1 \times P_2}}{P_{tot}}, \qquad (S8)$$

where $P_{tot} = P_1 + P_2$. For the powers used to record the spectra of Fig.4 in the main text ($P_1 = 27.5$mW, $P_2 = 6$mW) we obtain $m = 0.77$.

An accurate model of the QWIP detector should include the electrical capacitance of the PARs array, $C_{PAR}$, in parallel with $G(t)$. The Schottky contact biased in reverse breakdown should also be added (the forward biased Schottky junction is considered as a short circuit). The resulting electrical circuit model is shown in Fig. S5. This circuit also includes the inductor and capacitor ($L_{bias-T}$, $C_{bias-T}$) of the bias-T used in the experiment, allowing the decoupling between *dc* and *ac* currents. Finally, $Z_L$ is the load impedance seen by the QWIP (see Section 4). Voltages and currents in the circuit are time periodic (period $T=2\pi/\omega_b$) and can be expressed in a Fourier series. By neglecting high orders harmonics [3], the voltage across the photoconductor takes the simple form:

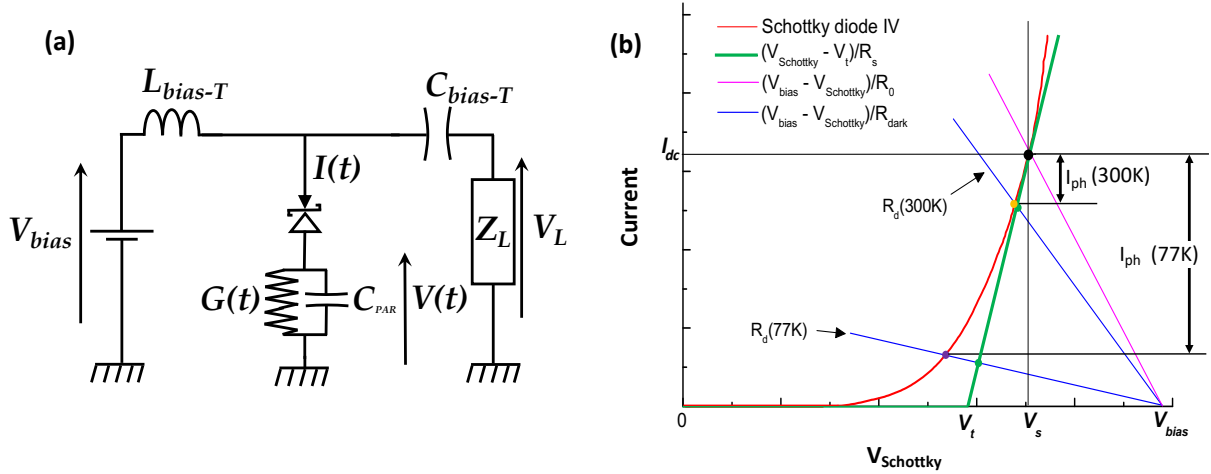

**Fig. S5 (a)** Electrical circuit model of the heterodyne mixing experiment. **(b)** Schematic Schottky diode IV characteristic in reverse breakdown (red). Modelled Schottky diode IV characteristic (green). Load lines under illumination (pink) and in the dark (blue), obtained from the circuit of Fig.S6(b). Operating point under illumination (black circle) and in the dark at 300K (orange circle) and 77K (purple circle). For simplicity, but without loss of generality, we have neglected the temperature dependence of the Schottky IV characteristic.

$$V(t) = V_{dc} + V_{ac}\cos(\omega_b t + \varphi). \qquad (S9)$$

The current is given by:

$$I(t) = I_{dc} + I_{ac}\cos(\omega_b t + \delta). \qquad (S10)$$

The quantities $V_{dc}, V_{ac}, I_{dc}, I_{ac}, \varphi, \delta$ can be derived from the circuit of Fig.S5 by applying Kirchhoff's laws at $\omega = 0$ and $\omega = \omega_b$ (i.e. exploiting the decoupling between *dc* and *ac* currents thanks to the bias-T inductance and capacitance) and by using the constitutive relation:

$$I(t) = G(t)V(t) \qquad (S11).$$

From the equations above, an *ac* small-signal circuit ($\omega = \omega_b$) and a *dc* circuit ($\omega = 0$) can be derived, as shown in Fig.S6(a),(b) [3]. Here, $R_S$ is the Schottky junction differential resistance *under illumination*

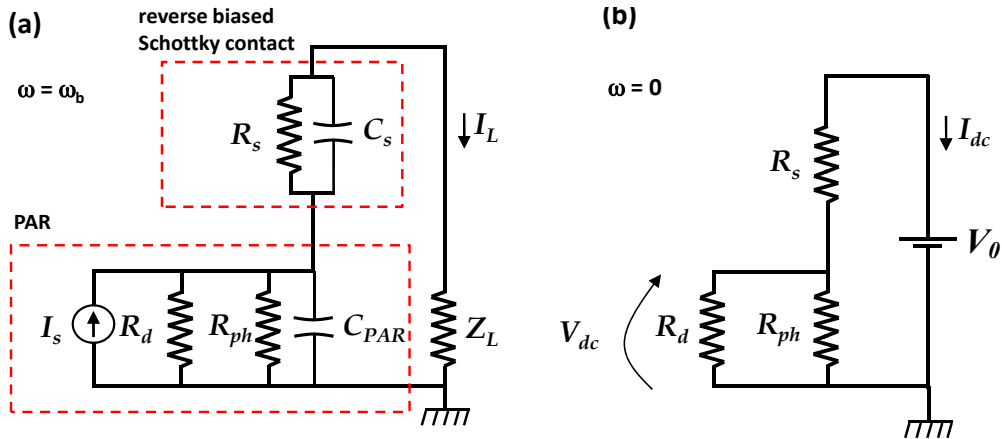

**Fig. S6 (a)** Equivalent small-signal *ac*

at the operating point $(I_{dc}, V_s)$: $1/R_S = dI/dV|_{I_{dc},V_s}$. In the *ac* circuit the small-signal behavior of the Schottky junction is modelled by adding a junction capacitance $C_s$ in parallel to $R_S$. Instead, in the *dc* circuit we must model the full Schottky junction IV curve. As shown in Fig.5(b) (green lines) this is done by replacing the latter (red line) by a linear characteristic $I = (V - V_t)/R_s$ passing through $(I_{dc}, V_s)$ for $V > V_t$, while for $V < V_t$ we consider the Schottky as an open circuit ($I = 0$). The threshold voltage $V_t$ can then be incorporated in the voltage source of the *dc* circuit by writing that $V_0 = (V_{bias} - V_t)$ for $V_{bias} > V_t$, and $V_0 = 0$ for $V_{bias} < V_t$ ($V_{bias}$ is the bias effectively applied to the device, see Fig.S5(a)). Clearly, both $V_t$ and $R_s$ will depend on the value of the QWIP point of operation under illumination.

In the *dc* circuit the PAR array is modeled by its *dc* photoresistance under illumination,

$$R_0 = \frac{R_d R_{ph}}{R_d + R_{ph}}, \qquad (S12)$$

where $R_d$ is the PAR array dark resistance. The corresponding load line (purple) is shown in Fig.5(b), with a slope given by $1/R_0$. The blue lines are instead the load lines of the PARs array *in the dark* at 300K and 77K, with slopes given by $1/R_d$.

At 300K the QWIP current under illumination is dominated by the dark current component, as can be seen from Fig.3(b) in the main text, and from Fig.S7(b),(d). In other words $R_d \sim R_0$, and the change of slope of the load line from illumination to dark (Fig.S5(b)) is small, i.e. the linearization of the Schottky IV through $R_s$ is a good approximation.

At 77K, the QWIP current under illumination is instead dominated by the photocurrent component (Fig.3(a) in the main text and Fig.S7(a),(c)), i.e. $R_d \gg R_0$, $R_s$, resulting into a large change of slope of the load line (Fig.S5(b)). In this case, by linearizing the Schottky using its resistance under illumination, we completely neglect the fact that in the dark the slope of the Schottky IV is much larger. On the other hand, as shown schematically in Fig.S5(b), the larger is $R_d$ the smaller will be the slope of the load line, thus reducing the difference between the effective dark current and the dark current obtained through the *dc* circuit model (purple and green circles).

In the *ac* circuit (Fig.S6(a)) the PAR array is modeled by an equivalent *ac* current source, $I_s$, with its internal impedance $R_0$ in parallel with the intrinsic capacitance of the array $C_{PAR}$. The current source can be computed as [3]:

$$I_s = V_{dc} \times G_1 = \frac{m}{\sqrt{1+(\omega_b \tau)^2}} V_{dc} \times \frac{1}{R_{ph}}, \qquad (S13)$$

where $I_s$ is in general a phasor (from now on we assume that all currents and voltages are represented by phasors). The *dc* equivalent circuit (Fig.S6(b)), can be used to derive $V_{dc}$:

$$V_{dc} = \frac{V_0}{R_0 + R_s} R_0 = I_{dc} R_0. \qquad (S14)$$

From this equation we note that, due to the Schottky contact resistance, $V_{dc} < V_0$ (note that the *effective dc* voltage applied to the PAR array is equal to $V_{dc} + V_t$). From Eqs.(S12), (S13) and (S14) we obtain:

$$I_s = \frac{m}{\sqrt{1+(\omega_b \tau)^2}} I_{dc} \frac{R_0}{R_{ph}} = \frac{m}{\sqrt{1+(\omega_b \tau)^2}} I_{dc} \frac{R_d - R_0}{R_d}, \qquad (S15)$$

where $I_{dc}$ is the *dc* current under illumination that can be measured experimentally. It is also useful to express the current source $I_s$ as a function of the *dc* photocurrent of the QWIP, which is obtained by subtracting the dark current from $I_{dc}$. From the *dc* equivalent circuit of Fig.S6(b) the photocurrent is given by:

$$I_{ph} = I_{dc} - I_{dark} = V_0\left[\frac{1}{R_0+R_s} - \frac{1}{R_d+R_s}\right] = I_{dc}\frac{R_d-R_0}{R_d+R_s} \quad (S16)$$

By comparing Eq.(S15) and (S16) we finally obtain:

$$I_s = \frac{m}{\sqrt{1+(\omega_b\tau)^2}}I_{ph}\frac{R_d+R_s}{R_d}. \quad (S17)$$

From Eq.(S16) we find that the *dc* photocurrent is equal to the *measured* dc current only if $R_0$ and $R_S$ are negligible compared to $R_d$. As can be deduced from Fig.3(a)(b) in the main text, this is the case at 77K for sufficiently high power levels, but not at 300K. In this case the correction factor $(R_d + R_s)/R_d$ (Eq.S17) cannot be neglected.

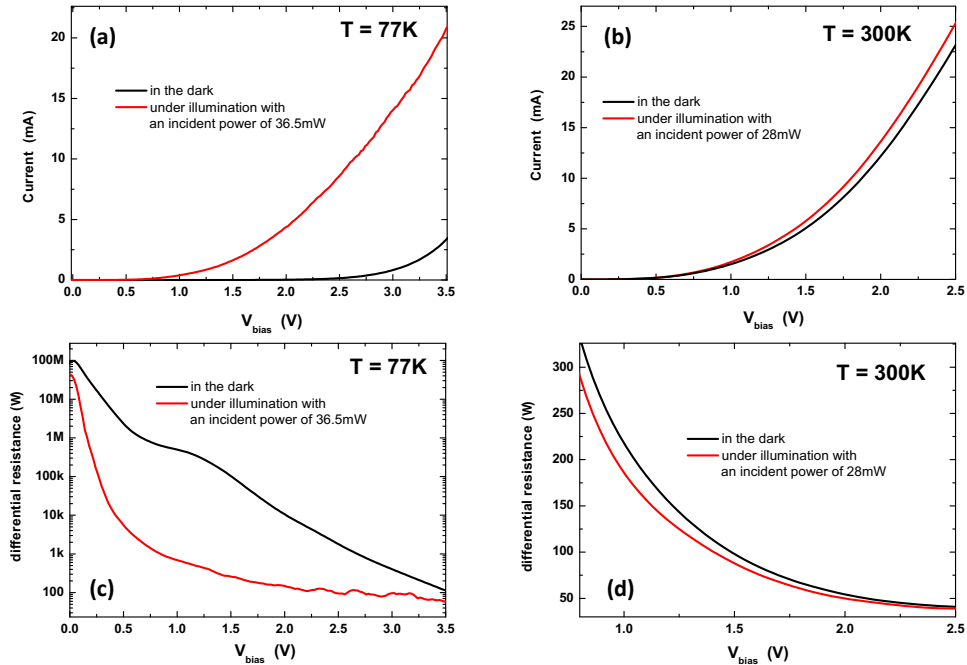

**Fig. S7 (a),(b)** Current under illumination and in the dark at 77K and 300K. **(c),(d)** Differential resistance under illumination and in the dark at 77K and 300K.

## 4. QWIP impedance measurements

In Fig.S8 we report the real and imaginary parts of the QWIP impedances *vs* frequency (black and red lines) obtained from the $S_{11}$ parameters measured with a VNA analyzer, after de-embedding the 50 integrated coplanar line. At T=77K, the $S_{11}$ parameters were measured under the same operating conditions (bias, temperature and illumination) used to record the FRs, while at 300K they were measured in the dark. This last choice stems from the fact that, contrary to 77K, at 300K the dark current is much larger than the photocurrent even under illumination at high power (see Fig. 3(b) of the main text), i.e. the QWIP impedance under illumination is very well approximated by the dark impedance ($R_d \ll R_{ph}$, see Fig.S6(a)).

The impedances at low biases (Fig.S8(a),(b)) are well reproduced by the equivalent circuit of Fig.S6(a), where the QWIP impedance (blue lines) is given by the sum of the PAR array and Schottky contact impedances:

$$Z_{QWIP}(\omega_b) = \frac{R_0}{1+i\omega_b R_0 C_{PAR}} + \frac{R_s}{1+i\omega_b R_s C_s} . \quad (S18)$$

In Fig.S8(a),(b), $Z_{QWIP}(\omega)$ is computed using the values of $R_0$, $R_s$- reported in the first and second column of Table 1 in the main text, with $C_{PAR} = 30$fF and $C_s$=0.7pF (see next Section). In particular, when $f \to 0$, we see clearly the effect of $C_s$, producing a fast increase of the real part of $Z_{QWIP}(\omega_b)$, until, at $f_b = 0$, $Re[Z_{QWIP}(\omega_b = 0)] = R_0 + R_s$. At higher frequencies ($f_b \gg (2\pi R_s C_s)^{-1}$), $R_s$ is shunted by $C_s$, and $Z_{QWIP}(\omega_b)$ coincides with the impedance of the PAR array, with a roll-off corresponding to a time constant equal to $R_0 C_{PAR}$ (first term in Eq.S18).

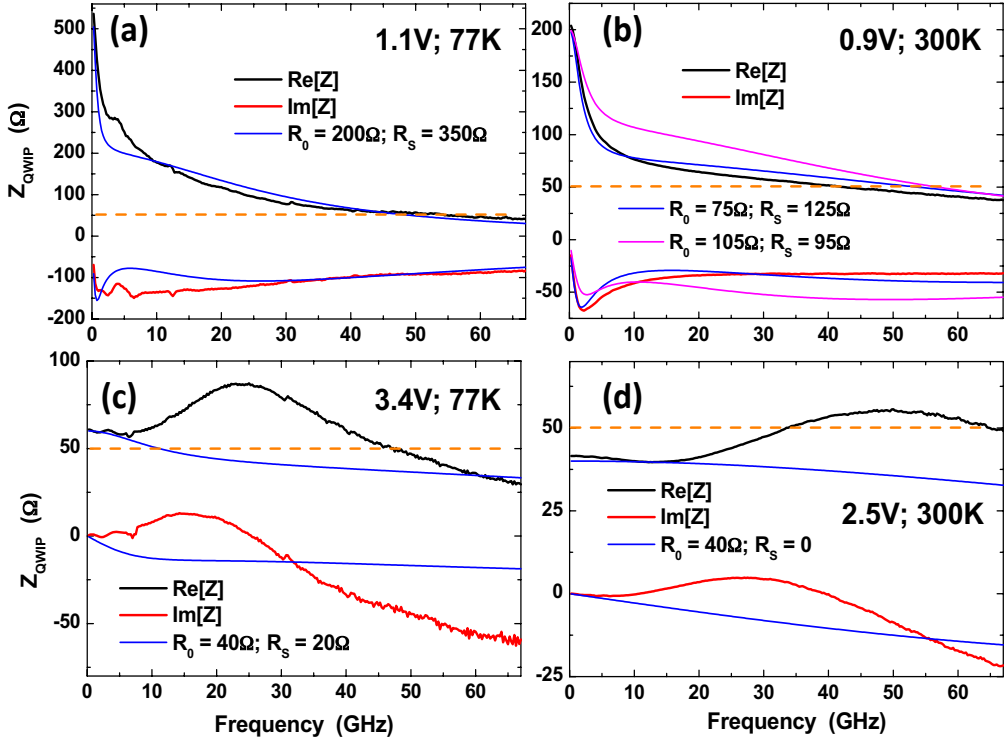

**Fig.S8**. Real (black) and imaginary (red) parts of the QWIP impedance, extracted from the $S_{11}$ parameters measurements, after de-embedding the 50Ω integrated coplanar line. The measurements at 77K (panels (a), (c)) were recorded under illumination with a power $P_{tot} = P_1 + P_2 = 33.5$mW. The measurements at 300K (panels (b),(d)) were done in the dark. The blue and purple lines (see text) represent the impedance computed from the small-signal equivalent circuit of Fig.S6(a) using the values of the resistances shown in the legends, with $C_{PAR} = 30$fF and $C_s$=0.7pF.

As shown in Fig.S8(c),(d), at high biases the QWIP impedances change completely. Firstly, the fast increase as $f_b \to 0$, disappears, which we interpret as the evidence that the Schottky junction becomes more transparent, i.e. $R_s$ shunts $C_s$ at all frequencies (see the next Section). At higher frequencies both the real and imaginary parts of $Z_{QWIP}(\omega_b)$ show a maximum, followed by a slow decay. As shown by the blue lines this behavior cannot be fully reproduced by our simple circuit model using the parameters reported in the third and fourth column of Table 1 of the main text. In particular the imaginary part becomes inductive around 15-30 GHz. This phenomenon is probably linked to the fact that the QWIP is operated close to the onset of intervalley scattering. A more detailed analysis is needed, which is beyond the scope of this work.

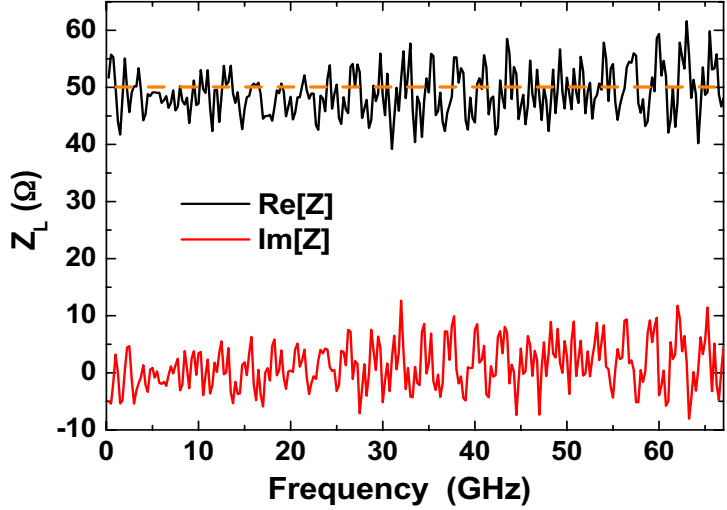

**Fig.S9.** Real and imaginary part of the impedance seen by the QWIP in the plane of the coplanar probes ($Z_L$).

In Fig.S9 we report the measured load impedance, $Z_L$, i.e. the impedance seen by the QWIP in the plane of the coplanar probes. This was extracted from $S_{11}$ parameter measurements. As can be seen, $Z_L$ can be approximated by its real part $Re[Z_L] = R_L \cong 50\Omega$.

## 5. Determination of QWIP circuit parameters

The various elements, $R_0, R_s, C_{PAR}$ and $C_s$ in the circuit of Fig.S6 depend in principle on the QWIP operating temperature, bias, and illumination conditions. To determine their values we rely on the experimental FR spectra displayed in Fig.4 of the main text and on the corresponding QWIP impedances shown in Fig.S8.

The first equation used to determine $R_0, R_s$, is given by (see Fig.S6):

$$R_0 + R_s = Re[Z_{QWIP}(\omega_b = 0)]. \qquad (S19)$$

The second equation is obtained instead by noting that the values of the experimental FRs in Fig.4 of the main text, correspond to the power dissipated in $Z_L$:

$$P_L = \tfrac{1}{2}\mathcal{R}e[Z_L] \cdot |I_L|^2 = \tfrac{1}{2}R_L \cdot |I_L|^2 . \quad (S20),$$

Indeed, we recall that the FRs are corrected by the attenuation from the QWIP to the SA, which, in turn, has an impedance of $50\Omega$, i.e. perfectly adapted to $Z_L$ (Fig. S8). Since $P_L$ in Eq.(S20) depends on $R_0$ and $R_s$, by comparing it with the power levels in the FRs of Fig.4, gives the second equation, which, together with Eq.(S19), allows the determination of the QWIP and Schottky resistances separately.

Concerning the capacitances, we begin by fixing $C_{PAR}$ = 28fF, which corresponds to the computed static capacitance of the PAR array using a parallel plate model. $C_s$ is instead determined by fitting the decay of the experimental FRs at low frequency.

The details of the calculations are given below, respectively at low bias (V_bias = 1.1V, 77K, and V_bias = 0.9V, 300K) and high bias (V_bias = 3.4V, 77K, and V_bias = 2.5V, 300K). The values of the measured *dc* photocurrent of the QWIP, $I_{ph}$, and the values of $R_0$ and $R_s$ can be found in Table 1 of the main text.

**Low bias**

We start by assuming to be at a sufficiently high frequency such that $R_S$ is shunted by the parallel capacitance $C_s$ ($f_b \gg (2\pi R_s C_s)^{-1}$) and can therefore be neglected (see Eq.(S18)). In this case we obtain:

$$I_L = I_s \frac{1}{1+R_L/R_0+i\omega_b R_L C_{PAR}}, \quad (S21)$$

where we have approximated $Z_L$ with its real part $R_L \cong 50\Omega$ (see Fig.S9). Now, provided that the frequency is not too high, e.g. $f_b \approx 10\text{GHz}$, the last term at the denominator can also be neglected thanks to the extremely low value of $C_{PAR}$ (note: the validity of these last two assumptions can be verified *a posteriori* from the values of $R_0$ and $R_s$). The power dissipated into the load is then given by:

$$P_L = \frac{1}{2} I_s^2 R_L \left[\frac{R_0}{R_0+R_L}\right]^2. \quad (S22)$$

At T=77K we have that $R_d \gg R_s$ (see Section 3), hence, from Eq.(S17), we have that $I_s \approx m \times I_{ph} = 0.38\text{mA}$ (at $f_b \approx 10\text{GHz}$ $\omega_b\tau$~0). At this point Eq.(S22) can be used to determine the value of $R_0$ by comparing $P_L$ with the measured value of the FR at 10GHz (1.1V, 77K curve in Fig4 of the main text ). The value of $R_s$ can finally be obtained from Eq.(S19) with $Re[Z_{QWIP}(\omega_b = 0)] = 550\Omega$ (see Fig.S8(a)). We find $R_0 = 200\Omega$ and $R_S = 350\Omega$ (first column of Table.1).

The last step consists in determining the value of $C_s$. This is obtained by fitting the decay of the experimental FR at low frequency (see Fig.4 in the main text), yielding $C_s \approx 0.7pF$. We note that this value is in agreement with the theoretical capacitance expected for a Au/GaAs Schottky junction with a doping density of 4x10$^{18}$ cm$^{-3}$ (~15nm depletion region width) [4]. The computed QWIP impedance is represented by the blue curves in Fig.S6(a), showing a good agreement with the impedance derived from the $S_{11}$ parameter. Also, the computed FR using Eq.(S19) reproduces very well the experimental one as shown in Fig.4 for $\tau$~1ps.

Concerning the measurement at T=300K and 0.9V, we have that $R_d \approx R_0$. In this case, from Eq.(S17) and Eq.(S22) we obtain:

$$P_L = \frac{1}{2} m^2 I_{ph}^2 \left[\frac{R_0+R_s}{R_0}\right]^2 \left[\frac{R_0}{R_0+R_L}\right]^2 R_L = \frac{1}{2} m^2 I_{ph}^2 \left[\frac{R_0+R_s}{R_0+R_L}\right]^2 R_L \quad (S23)$$

where, again, we used the fact that at $f_b \approx 10\text{GHz}$ $\omega_b\tau$~0. In this last equation the term $R_0 + R_s$ is known from Eq.(S19) and Fig.S8(b) ($R_0 + R_s = Re[Z_{QWIP}(\omega_b = 0)] = 200\Omega$). Then, again, $R_0$ is determined by comparing $P_L$ in Eq.(S23) with the measured value of the FR at 10GHz (0.9V, 300K curve in Fig4 of the main text). From this procedure we obtain $R_0 = 105\Omega$ and $R_S = 95\Omega$, which, however, do not allow to reproduce the QWIP impedance in a satisfactory way, as shown by the purple traces in Fig.S8(b) (here we used $C_s = 0.7\text{pF}$). We find that the values $R_0 = 75\Omega$ and $R_S = 125\Omega$, allow to obtain the closest agreement with $P_L$, compatibly with a good fit of the QWIP impedance (blue traces in Fig.S8(b)). The resulting computed FR, shown in Fig.4 of the main text, is ~2dBm above the measured FR. This spectrum was obtained with $C_s = 0.7\text{pF}$, yielding, as for the 77K,1.1V FR, a decay at low frequency in good agreement with the measurement.

**High bias**

As already pointed out, at high bias we don't observe anymore the drop in the FR as $f_b \to 0$. In other words, $C_s$ is shunted by $R_s$, which can be taken as the Schottky contact impedance at virtually all frequencies. As we did at low bias, we also assume that the frequency is sufficiently low that $2\pi f_b R_0 C_{PAR} \ll 1$ (e.g. $f_b$ = 1GHz). Under these assumptions we have that with $I_L = I_s \times R_0/(R_0 + R_S + R_L)$, yielding:

$$P_L = \frac{1}{2} I_s^2 R_L \left[\frac{R_0}{R_0 + R_S + R_L}\right]^2, \quad \text{(S24)}$$

where, as usual, $R_0 + R_s = Re[Z_{QWIP}(\omega = 0)]$.

At T=77K, since $R_d \gg R_s$, we have that $I_s \approx m \times I_{ph} = 11.7\text{mA}$, and $R_0 + R_s = 60\Omega$ (see Fig.S8(c)). By using $P_L$ from Eq.(S24) to fit the value of the measured FR at 1GHz (3.4V, 77K curve in Fig4 of the main text) we obtain $R_0 = 40\Omega$ and $R_s = 20\Omega$. As shown in Fig.4 of the main text, from Eq.(S20) we obtain an excellent agreement with the measured FR using $\tau$ ~8ps.

At T=300K we still have $R_d \ll R_{ph}$, i.e. $R_d \approx R_0$. Hence, from Eq.(S17) and (S24) we have:

$$P_L = \frac{1}{2} m^2 I_{ph}^2 \left[\frac{R_0 + R_S}{R_0 + R_S + R_L}\right]^2 R_L, \quad \text{(S25)}$$

with $R_0 + R_s = Re[Z_{QWIP}(\omega_b = 0)] = 40\Omega$ (see Fig.S8(d)). Since Eq.(S25) also depends on the sum $R_0 + R_s$, in this case the values of $R_0$ and $R_S$ are determined by fitting the measured FR over the full frequency range using Eq.(S20) (with $m \times I_{ph} = 1.7\text{mA}$). The best agreement is obtained with $R_0 = 40\Omega$ and $R_s = 0$ (Fig. 4 in the main text). As was done at low bias, by subtracting $R_s$ from the QWIP differential resistance in the dark (Fig.S7(d)), we obtain the value of the PAR array dark resistance at 2.5V (300K): $R_d$ ~ 42$\Omega$.

As shown in Fig.S8(c),(d), contrary to what happens at low bias, the computed impedances at high bias provide only an approximated value of the actual QWIP impedance (see Section 4).

## 6. Evaluation of carriers capture and transit times

**T = 77K, V_bias = 3.4V.**

From the responsivity reported in Fig.3(c) of the main text at $P_{tot} = P_1 + P_2$=33.5mW, we obtain $g(77K, 3.4V) = \tau_c/\tau_{tr} \simeq 1.25$ (i.e. ~ half the value at low incident power $g \simeq 2.5$ – see Section 1). The roll-off time constant $\tau$ can therefore be approximated by the transit time [5]. From the fit of Fig.4 in the main text, we then have that $\tau \simeq \tau_{tr} \simeq 8\text{ps}$ and $\tau_c = 1.25 \times \tau_{tr} \simeq 10\text{ps}$

**T = 77K, V_bias = 1.1V.**

We have that $I_{ph} \propto g$, therefore (see Table 1 in the main text):

$$g(77K, 1.1V) = g(77K, 3.4V) \times \frac{I_{ph}(77K,1.1V)}{I_{ph}(77K,3.4V)} \simeq 1.25 \times \frac{0.49mA}{15.2mA} = 0.04. \quad \text{(S26)}$$

The roll-off time constant $\tau$ can therefore be approximated by the capture time. From the fit of Fig.4 in the main text, we then have that $\tau \simeq \tau_c \simeq 1\text{ps}$, and $\tau_{tr} = \tau_c/0.04 \simeq 25\text{ps}$.

**T = 300K, V$_{bias}$ = 0.9V.**

Following the same procedure described above we obtain $g(300K, 0.9V) \simeq 0.011$. From the fit of Fig.4 in the main text, we then have that $\tau \simeq \tau_c \lesssim 1\text{ps}$, and $\tau_{tr} \gtrsim \tau_c/0.011 \simeq 90\text{ps}$.

**T = 300K, V$_{bias}$ = 2.5V.**

Following the same procedure described above we obtain $g(300K, 2.5V) \simeq 0.18$. From the fit of Fig.4 in the main text, we then have that $\tau \simeq \tau_c \simeq 2.5\text{ps}$, and $\tau_{tr} \simeq \tau_c/0.18 \simeq 14\text{ps}$.

**References**

1. Y. N. Chen, Y. Todorov, B. Askenazi, A. Vasanelli, G. Biasiol, R. Colombelli, and C. Sirtori *Antenna-coupled microcavities for enhanced infrared photo-detection*, Appl. Phys. Lett. **104**, 031113-1 (2014).
2. T. Ando, A. Fowler, and F. Stern, *Electronic properties of two dimensional systems*, Rev. Mod. Phys. 54, 437 (1982).
3. P. D. Coleman, R. C. Eden, and J. N. Weaver, *Mixing and Detection of Coherent Light*, IEEE Trans. Electron Devices, 11, 488 (1964)
4. S. M. Sze and K. K. Ng in Physics of Semiconductor Devices (Wiley, 2006)
5. Schneider, H. and Liu, H. C. in Quantum Well Infrared Photodetectors: Physics and Applications 72–75 (Springer, 2007).